# Weight Distribution of A $p$-ary Cyclic Code

Xiangyong Zeng[1], Lei Hu[2], Wenfeng Jiang[2], Qin Yue[3], Xiwang Cao[3]

1. Faculty of Mathematics and Computer Science, Hubei University, Wuhan, 430062, China

E-mail: xzeng@hubu.edu.cn

2. State Key Laboratory of Information Security, Graduate School of

Chinese Academy of Sciences, Beijing, 100049, China

E-mail:{hu,wfjiang}@is.ac.cn

3. Department of Math, School of Sciences

Nanjing University of Aeronautics and Astronautics, Nanjing 210016, China

E-mail:{yueqin,xwcao}@nuaa.edu.cn

**Abstract:** For an odd prime $p$ and two positive integers $n \geq 3$ and $k$ with $\frac{n}{\gcd(n,k)}$ being odd, the paper determines the weight distribution of a $p$-ary cyclic code $\mathcal{C}$ over $\mathbb{F}_p$ with nonzeros $\alpha^{-1}$, $\alpha^{-(p^k+1)}$ and $\alpha^{-(p^{3k}+1)}$, where $\alpha$ is a primitive element of $\mathbb{F}_{p^n}$.

## 1 Introduction

Nonlinear functions over finite fields have useful applications in coding theory and cryptography [2, 15]. Some linear codes having good properties [3, 5, 7, 13, 15, 17] were constructed from highly nonlinear functions [4, 6, 8, 16].

Let $q$ be a power of a prime $p$, and $\mathbb{F}_{q^n}$ be a finite field with $q^n$ elements. A $p$-ary $[m, l]$ linear code is a linear subspace of $\mathbb{F}_p^m$ with dimension $l$. The Hamming weight of a codeword $c_1 c_2 \cdots c_m$ is the number of nonzero $c_i$ for $1 \leq i \leq m$. In this paper, we study a $[p^n - 1, 3n]$ cyclic code $\mathcal{C}$ given by

$$\mathcal{C} = \left\{ \mathbf{c}(\epsilon, \gamma, \delta) = \left( \mathrm{Tr}_1^n(\epsilon x + \gamma x^{p^k+1} + \delta x^{p^{3k}+1}) \right)_{x \in \mathbb{F}_{p^n}^*} \mid \epsilon, \gamma, \delta \in \mathbb{F}_{p^n} \right\},$$

where $k$ is a positive integer and $\mathrm{Tr}_1^n$ is the trace function from $\mathbb{F}_{p^n}$ to $\mathbb{F}_p$. This code is constructed from the function $\mathrm{Tr}_1^n(\epsilon x + \gamma x^{p^k+1} + \delta x^{p^{3k}+1})$, which can have high nonlinearity if either $\gamma$ or $\delta$ is nonzero. It is easy to know that $\alpha^{-1}$, $\alpha^{-(p^k+1)}$, $\alpha^{-(p^{3k}+1)}$ and their $\mathbb{F}_p$-conjugates are all nonzeros of the cyclic code $\mathcal{C}$, where $\alpha$ is a primitive element of $\mathbb{F}_{p^n}$ [15].

In this paper we assume that $p$ and $\frac{n}{\gcd(k,n)}$ are both odd, and we determine the weight distribution of the code $\mathcal{C}$. To this goal, we will focus on determining the ranks of a class of quadratic forms and calculating two classes of exponential sums. The ranks of quadratic forms are determined through finding the number of solutions to a class of linearized polynomials

$$L_{\gamma,\delta}(z) = \gamma z^{p^k} + \gamma^{p^{-k}} z^{p^{-k}} + \delta z^{p^{3k}} + \delta^{p^{-3k}} z^{p^{-3k}}$$

over the field $\mathbb{F}_{p^n}$. By applying the theory of quadratic forms, two classes of exponential sums are evaluated and the weight distribution of the cyclic code $\mathcal{C}$ is determined. Throughout the paper, we set $d = \gcd(k, n)$, $s = \frac{n}{\gcd(k,n)}$ and $n \geq 3$.

The remainder of this paper is organized as follows. Section 2 gives some definitions and preliminaries. Section 3 studies the rank distribution of a class of quadratic forms. Section 4 determines the weight distribution of $\mathcal{C}$.

## 2 Preliminaries

Identifying $\mathbb{F}_{q^n}$ with the $n$-dimensional $\mathbb{F}_q$-vector space $\mathbb{F}_q^n$, a univariable polynomial $f(x)$ defined on $\mathbb{F}_{q^n}$ can be regarded as an $n$-variable polynomial on $\mathbb{F}_q$. The former is called a quadratic form if the latter is a homogeneous polynomial of degree two:

$$f(x_1, \cdots, x_n) = \sum_{1 \leq j \leq k \leq n} a_{jk} x_j x_k,$$

here we use a basis of $\mathbb{F}_q^n$ over $\mathbb{F}_q$ and identify $x \in \mathbb{F}_{q^n}$ with a vector $(x_1, \cdots, x_n) \in \mathbb{F}_q^n$. The rank of the quadratic form $f(x)$ is defined as the codimension of the $\mathbb{F}_q$-vector space

$$W = \{w \in \mathbb{F}_{q^n} \mid f(x+w) = f(x) \text{ for all } x \in \mathbb{F}_{q^n}\}, \tag{1}$$

denoted by $\mathrm{rank}(f)$. Then $|W| = q^{n-\mathrm{rank}(f)}$.

For a quadratic form $f(x)$, there exists a symmetric matrix $A$ such that $f(x) = X^{\mathrm{T}} A X$, where $X$ is written as a column vector and its transpose is $X^{\mathrm{T}} = (x_1, x_2, \cdots, x_n) \in \mathbb{F}_q^n$. The determinant $\det(f)$ of $f(x)$ is defined to be the determinant of $A$, and $f(x)$ is nondegenerate if $\det(f) \neq 0$. By Theorem 6.21 of [14], there exists a nonsingular matrix $B$ such that $B^{\mathrm{T}} A B$ is a diagonal matrix. Making a nonsingular linear substitution $X = BY$ with $Y^{\mathrm{T}} = (y_1, y_2, \cdots, y_n)$, one has

$$f(x) = Y^{\mathrm{T}} B^{\mathrm{T}} A B Y = \sum_{i=1}^{r} a_i y_i^2 \tag{2}$$

where $r \leq n$ is the rank of $f(x)$ and $a_1, a_2, \cdots, a_r \in \mathbb{F}_q^*$.

Let $m$ be a positive factor of the integer $n$. The trace function $\mathrm{Tr}_m^n$ from $\mathbb{F}_{p^n}$ to $\mathbb{F}_{p^m}$ is defined by

$$\mathrm{Tr}_m^n(x) = \sum_{i=0}^{n/m-1} x^{p^{mi}}, \quad x \in \mathbb{F}_{p^n}.$$

Let $e(y) = e^{2\pi\sqrt{-1}\, \mathrm{Tr}_1^n(y)/p}$ and $\zeta_p = e^{2\pi\sqrt{-1}/p}$.

The following lemmas will be used throughout this paper.

**Lemma 1 (Theorems 5.33 and 5.15 of [14]):** Let $\mathbb{F}_q$ be a finite field with $q = p^l$, where $p$ is an odd prime. Let $\eta$ be the quadratic character of $\mathbb{F}_q$. Then for $a \neq 0$,

$$\sum_{x \in \mathbb{F}_q} \zeta_p^{\mathrm{Tr}_1^l(ax^2)} = \begin{cases} \eta(a)(-1)^{l-1} p^{\frac{l}{2}}, & \text{if } p \equiv 1 \,(\mathrm{mod}\, 4), \\ \eta(a)(-1)^{l-1} (\sqrt{-1})^l p^{\frac{l}{2}}, & \text{if } p \equiv 3 \,(\mathrm{mod}\, 4). \end{cases}$$

**Lemma 2 (Theorems 6.26 and 6.27 of [14]):** Let $q$ be an odd prime power, $f$ be a nondegenerate quadratic form over $\mathbb{F}_{q^l}$. Define a function $v(\rho)$ over $\mathbb{F}_q$ as $v(0) = q - 1$ and $v(\rho) = -1$ for $\rho \in \mathbb{F}_q^*$. Then for $\rho \in \mathbb{F}_q$ the number of solutions to the equation $f(x_1, \cdots, x_l) = \rho$ is

$$q^{l-1} + q^{\frac{l-1}{2}} \eta\left((-1)^{\frac{l-1}{2}} \rho \cdot \det(f)\right)$$

for odd $l$, and

$$q^{l-1} + v(\rho) q^{\frac{l-2}{2}} \eta\left((-1)^{\frac{l}{2}} \det(f)\right)$$

for even $l$.

**Lemma 3:** (i) (**Theorems 5.15 of [14]**) Let $\zeta_p$ be a complex primitive $p$-th root of unity and $\eta$ be the quadratic character of $\mathbb{F}_p$. Then

$$\sum_{\rho=1}^{p-1} \eta(\rho) \zeta_p^\rho = \sqrt{(-1)^{\frac{p-1}{2}} p}.$$

(ii) Let the function $v(\rho)$ over $\mathbb{F}_p$ be defined by $v(0) = p - 1$ and $v(\rho) = -1$ for $\rho \in \mathbb{F}_p^*$. Then

$$\sum_{\rho=1}^{p-1} v(\rho)\zeta_p^\rho = p.$$

## 3 Rank distribution of a class of quadratic forms

This section studies the rank distribution of the quadratic forms $\mathrm{Tr}_d^n(\gamma x^{p^k+1} + \delta x^{p^{3k}+1})$ for nonzero $\gamma$ or $\delta$.

To determine the distribution, we define a related exponential sum

$$S(\epsilon, \gamma, \delta) = \sum_{x \in \mathbb{F}_{p^n}} e\left(\epsilon x + \gamma x^{p^k+1} + \delta x^{p^{3k}+1}\right), \quad \epsilon,\, \gamma,\, \delta \in \mathbb{F}_{p^n}. \tag{3}$$

Then the possible values of the ranks are measured by evaluating the exponential sum $S(\epsilon, \gamma, \delta)$. For discussion of the exponential sum of a general quadratic form, please refer to references [10] and [12].

**Proposition 1:** For odd $s$ and $\delta \in \mathbb{F}_{p^n}^*$, the exponential sum $S(\epsilon, \gamma, \delta)$ satisfies

$$|S(\epsilon, \gamma, \delta)| = 0,\ p^{\frac{n}{2}},\ p^{\frac{n+d}{2}},\ \text{or}\ p^{\frac{n}{2}+d}.$$

**Proof:** Notice that

$$\begin{aligned}
&|S(\epsilon, \gamma, \delta)|^2 \\
&= \overline{S(\epsilon, \gamma, \delta)}\, S(\epsilon, \gamma, \delta) \\
&= \sum_{x \in \mathbb{F}_{p^n}} e\left(-\epsilon x - \gamma x^{p^k+1} - \delta x^{p^{3k}+1}\right) \sum_{y \in \mathbb{F}_{p^n}} e\left(\epsilon y + \gamma y^{p^k+1} + \delta y^{p^{3k}+1}\right) \\
&= \sum_{x,\, z \in \mathbb{F}_{p^n}} e\left(\epsilon z + \gamma z^{p^k+1} + \delta z^{p^{3k}+1} + \gamma z^{p^k} x + \gamma z x^{p^k} + \delta z^{p^{3k}} x + \delta z x^{p^{3k}}\right) \quad \text{by } y = x + z \\
&= \sum_{z \in \mathbb{F}_{p^n}} e\left(\epsilon z + \gamma z^{p^k+1} + \delta z^{p^{3k}+1}\right) \sum_{x \in \mathbb{F}_{p^n}} e\left(x L_{\gamma, \delta}(z)\right)
\end{aligned} \tag{4}$$

where

$$L_{\gamma, \delta}(z) = \gamma z^{p^k} + \gamma^{p^{-k}} z^{p^{-k}} + \delta z^{p^{3k}} + \delta^{p^{-3k}} z^{p^{-3k}}$$

is a linearized polynomial in $z$. Let $V$ be the set of all roots of $L_{\gamma,\delta}(z) = 0$. (By abuse of notation, we use $V$ to denote the set in despite of it depending on $\gamma$ and $\delta$.) Thus, $V$ is an $\mathbb{F}_{p^d}$-vector space. By (4), we have

$$|S(\epsilon, \gamma, \delta)|^2 = p^n \left|\sum_{z \in V} e\left(\epsilon z + \gamma z^{p^k+1} + \delta z^{p^{3k}+1}\right)\right|. \tag{5}$$

Let

$$\Phi_{\gamma,\delta}(x) = \gamma x^{p^k+1} + \delta x^{p^{3k}+1} - \delta^{p^{-k}} x^{p^{2k}+p^{-k}} + \delta^{p^{-2k}} x^{p^k+p^{-2k}}. \tag{6}$$

By (6), we have

$$\mathrm{Tr}_1^n(\Phi_{\gamma,\delta}(z)) = \mathrm{Tr}_1^n\left(\gamma z^{p^k+1} + \delta z^{p^{3k}+1}\right) \tag{7}$$

and

$$\Phi_{\gamma,\delta}(z) + \Phi_{\gamma,\delta}(z)^{p^{-k}} = z L_{\gamma,\delta}(z). \tag{8}$$

If $z \in V$, then by (8),

$$\Phi_{\gamma,\delta}(z)^{p^k} = -\Phi_{\gamma,\delta}(z). \tag{9}$$

By $\gcd(k,n) = d$, there is an integer $k'$ such that $kk' \equiv d \pmod{n}$ and hence, $\Phi_{\gamma,\delta}(z)^{p^d} = \Phi_{\gamma,\delta}(z)^{p^{kk'}} = (-1)^{k'}\Phi_{\gamma,\delta}(z)$, where the last equality is derived from (9). If $k'$ is even, $\Phi_{\gamma,\delta}(z)^{p^d} = \Phi_{\gamma,\delta}(z)$ and then $\Phi_{\gamma,\delta}(z)^{p^k} = \Phi_{\gamma,\delta}(z)$, which together with (9) again implies $\Phi_{\gamma,\delta}(z) = 0$. If $k'$ is odd, then

$$\Phi_{\gamma,\delta}(z)^{p^d} = -\Phi_{\gamma,\delta}(z). \tag{10}$$

By the property $\text{Tr}_d^n(\Phi_{\gamma,\delta}(z)) = \text{Tr}_d^n(\Phi_{\gamma,\delta}(z)^{p^{-d}})$ of trace function and (8), we have

$$\begin{aligned} 0 &= \text{Tr}_d^n(zL_{\gamma,\delta}(z)) \\ &= \text{Tr}_d^n(\Phi_{\gamma,\delta}(z)) + \text{Tr}_d^n(\Phi_{\gamma,\delta}(z)^{p^{-k}}) \\ &= 2\text{Tr}_d^n(\Phi_{\gamma,\delta}(z)) \\ &= 2(\Phi_{\gamma,\delta}(z) + \Phi_{\gamma,\delta}(z)^{p^d} + \cdots + \Phi_{\gamma,\delta}(z)^{p^{(s-1)d}}) \\ &= 2\Phi_{\gamma,\delta}(z), \end{aligned}$$

where the last equal sign holds due to (10) and $s$ being odd. This implies $\Phi_{\gamma,\delta}(z) = 0$ and $\text{Tr}_1^n(\gamma z^{p^k+1} + \delta z^{p^{3k}+1}) = \text{Tr}_1^n(\Phi_{\gamma,\delta}(z)) = 0$ by (7). Conversely, if $\Phi_{\gamma,\delta}(z) = 0$, then by (8), $L_{\gamma,\delta}(z) = 0$ and $\text{Tr}_1^n(\Phi_{\gamma,\delta}(z)) = 0$. Therefore, $z \in V$ if and only if $\Phi_{\gamma,\delta}(z) = 0$. Further, in this case $\text{Tr}_1^n(\gamma z^{p^k+1} + \delta z^{p^{3k}+1}) = 0$. Thus, by (5),

$$|S(\epsilon, \gamma, \delta)| = \sqrt{p^n \left| \sum_{z \in V} \zeta_p^{\text{Tr}_1^n(\epsilon z)} \right|}. \tag{11}$$

Since $V$ is an $\mathbb{F}_{p^d}$-vector space, we can assume $|V| = p^{dm}$ for an integer $m \geq 0$.

If $m \geq 3$, then $\Phi_{\gamma,\delta}(z) = 0$ has at least $p^{3d}$ solutions. For a fixed $z_0 \in V\setminus\{0\}$ and for any $z \in V$, we have $\Phi_{\gamma,\delta}(z) = \Phi_{\gamma,\delta}(z_0) = 0$ and $\Phi_{\gamma,\delta}(z+z_0) = 0$ since $z+z_0$ is also in the vector space $V$. Thus, the equation

$$(z+z_0)(z_0\Phi_{\gamma,\delta}(z) + z\Phi_{\gamma,\delta}(z_0)) - zz_0\Phi_{\gamma,\delta}(z+z_0) = 0 \tag{12}$$

has at least $p^{3d}$ solutions.

By (6), Equation (12) becomes

$$\delta^{p^{-2k}}(z^{p^k}z_0 - zz_0^{p^k})(z^{p^{-2k}}z_0 - zz_0^{p^{-2k}}) - \delta^{p^{-k}}(z^{p^{2k}}z_0 - zz_0^{p^{2k}})(z^{p^{-k}}z_0 - zz_0^{p^{-k}}) = 0, \tag{13}$$

which has at least $p^{3d}$ roots on variable $z$. Let $z = wz_0$, then

$$\delta^{p^{-2k}} z_0^{p^k+p^{-2k}+2}(w^{p^k} - w)(w^{p^{-2k}} - w) - \delta^{p^{-k}} z_0^{p^{2k}+p^{-k}+2}(w^{p^{2k}} - w)(w^{p^{-k}} - w) = 0.$$

Let $u = w^{p^{-k}} - w$, the above equation can be rewritten as

$$-\delta^{p^{-2k}} z_0^{p^k+p^{-2k}} u^{p^k}(u^{p^{-k}} + u) + \delta^{p^{-k}} z_0^{p^{2k}+p^{-k}}(u^{p^{2k}} + u^{p^k})u = 0,$$

which has at least $p^{2d}$ roots on $u$ since $w^{p^{-k}} - w = u$ has at most $p^d$ roots on $w$ for each $u$. Define

$$\Psi_{\delta,z_0}(x) = \delta^{p^{-2k}} z_0^{p^k+p^{-2k}} x^{p^k}(x^{p^{-k}} + x) - \delta^{p^{-k}} z_0^{p^{2k}+p^{-k}}(x^{p^{2k}} + x^{p^k})x.$$

Similarly, for each nonzero root $u_0$ of $\Psi_{\delta,z_0}(u) = 0$, the equation

$$(u+u_0)(u_0\Psi_{\delta,z_0}(u) + u\Psi_{\delta,z_0}(u_0)) - uu_0\Psi_{z,z_0}(u+u_0) = 0$$

has at least $p^{2d}$ solutions on $u$. By the definition of $\Psi_{\delta,z_0}(x)$, the above equation is equivalent to

$$\delta^{p^{-2k}} z_0^{p^k+p^{-2k}}(u^{p^k}u_0 - uu_0^{p^k})(u^{p^{-k}}u_0 - uu_0^{p^{-k}}) = 0.$$

This shows that $u = vu_0$ where $v \in \mathbb{F}_{p^d}$. Consequently, for each given $u_0 \neq 0$, the above equation has at most $p^d$ roots. This gives a contradiction and then $m \leq 2$.

This finishes the proof. ∎

**Remark 1:** The possible ranks of some quadratic forms can be determined by directly calculating the number of the solutions to their related linearized polynomials [18, 11]. The number of the roots to the linearized polynomial $L_{\gamma,\delta}(z)$ in Proposition 1 is discussed by studying that of an associated nonlinear polynomial. The method was first presented to study a linear mapping over a finite field of characteristic 2 [9] and further used to discuss some triple error correcting binary codes with BCH parameters [1]. In Proposition 1, we used this method to the cases of odd characteristic.

From Proposition 1, the value of the dimension $m$ determines the rank of the following quadratic form.

**Corollary 1:** For odd $s$ and $\delta \in \mathbb{F}_{p^n}^*$, the quadratic form

$$\Omega_{\gamma,\delta}(x) = \mathrm{Tr}_d^n\left(\gamma x^{p^k+1} + \delta x^{p^{3k}+1}\right)$$

has rank $s$, $s-1$, or $s-2$.

When there is exactly one nonzero element in $\{\gamma, \delta\}$, the rank of $\Omega_{\gamma,\delta}(x)$ can be determined by directly calculating the number of solutions to $L_{\gamma,\delta}(z) = 0$.

**Proposition 2:** For odd $s$ and $\gamma, \delta \in \mathbb{F}_{p^n}^*$, the quadratic forms $\Omega_{\gamma,0}(x) = \mathrm{Tr}_d^n\left(\gamma x^{p^k+1}\right)$ and $\Omega_{0,\delta}(x) = \mathrm{Tr}_d^n\left(\delta x^{p^{3k}+1}\right)$ have rank $s$.

**Proof:** We only give the proof of $\mathrm{rank}(\Omega_{0,\delta}) = s$, and another case can be proven in a similar way.

It is sufficient to determine the number of solutions to $\delta z^{p^{3k}} + \delta^{p^{-3k}} z^{p^{-3k}} = 0$. This equation has nonzero solutions if and only if $(\delta z^{p^{3k}+1})^{p^{3k}-1} = -1$. If the latter holds, then $\gcd(p^{3k}-1, p^n-1) = (p^{\gcd(3k,n)}-1) | \frac{p^n-1}{2}$. Let $s_1 = \frac{n}{\gcd(3k,n)}$ and then

$$p^n - 1 = \left(p^{\gcd(3k,n)} - 1\right)\left(p^{(s_1-1)\gcd(3k,n)} + p^{(s_1-2)\gcd(3k,n)} + \cdots + p^{\gcd(3k,n)} + 1\right).$$

Notice that $s_1$ is a factor of the odd integer $s$. As a consequence, $p^{(s_1-1)\gcd(3k,n)} + p^{(s_1-2)\gcd(3k,n)} + \cdots + p^{\gcd(3k,n)} + 1$ is odd and $\frac{p^n-1}{2}$ can not be divided by $p^{\gcd(3k,n)} - 1$. Thus, $-1$ is not $(p^{\gcd(3k,n)} - 1)$-th power of any element in $\mathbb{F}_{p^n}^*$ and then $\delta z^{p^{3k}} + \delta^{p^{-3k}} z^{p^{-3k}} = 0$ has only the zero solution. This finishes the proof. ∎

**Remark 2:** For $\gamma, \delta \in \mathbb{F}_{p^n}^*$, $\mathrm{Tr}_1^d\left(\Omega_{\gamma,0}(x)\right)$ and $\mathrm{Tr}_1^d\left(\Omega_{0,\delta}(x)\right)$ are $p$-ary bent functions.

To study the rank distribution of the quadratic form $\Omega_{\gamma,\delta}$, for $i \in \{0,1,2\}$, we define

$$R_i = \left\{(\gamma,\delta) \mid \mathrm{rank}(\Omega_{\gamma,\delta}) = s - i, (\gamma,\delta) \in \mathbb{F}_{p^n} \times \mathbb{F}_{p^n} \setminus \{(0,0)\}\right\}. \tag{14}$$

**Lemma 4:** $|R_2| = \frac{(p^{n-d}-1)(p^n-1)}{p^{2d}-1}$.

**Proof:** If $(\gamma,\delta) \in R_2$, then $\gamma\delta \neq 0$ by Propositions 1 and 2, and $V$ is a two-dimensional vector space over $\mathbb{F}_{p^d}$. Let $\{v_1, v_0\}$ be a basis of $V$ over $\mathbb{F}_{p^d}$. Then, $v_1 v_0^{-1} \notin \mathbb{F}_{p^d}$ and $(v_1^{p^{4k}} v_0^{p^{2k}} - v_1^{p^{2k}} v_0^{p^{4k}})(v_1^{p^k} v_0^{p^{2k}} - v_1^{p^{2k}} v_0^{p^k}) \neq 0$. By (13),

$$\begin{aligned}
\delta^{p^k-1} &= \frac{(v_1^{p^{3k}} v_0^{p^{2k}} - v_1^{p^{2k}} v_0^{p^{3k}})(v_1 v_0^{p^{2k}} - v_1^{p^{2k}} v_0)}{(v_1^{p^{4k}} v_0^{p^{2k}} - v_1^{p^{2k}} v_0^{p^{4k}})(v_1^{p^k} v_0^{p^{2k}} - v_1^{p^{2k}} v_0^{p^k})} \\
&= \left(\frac{v_1^{p^{2k}} v_0^{p^k} - v_1^{p^k} v_0^{p^{2k}}}{(v_1^{p^{2k}} v_0 - v_1 v_0^{p^{2k}})^{p^k+1}}\right)^{p^k-1}.
\end{aligned}$$

Thus,

$$\delta = \lambda \frac{v_1^{p^{2k}} v_0^{p^k} - v_1^{p^k} v_0^{p^{2k}}}{(v_1^{p^{2k}} v_0 - v_1 v_0^{p^{2k}})^{p^k+1}} \tag{15}$$

for an element $\lambda \in \mathbb{F}_{p^d}^*$. Since $\Phi_{\gamma,\delta}(v_1) = \gamma v_1^{p^k+1} + \delta v_1^{p^{3k}+1} - \delta^{p^{-k}} v_1^{p^{2k}+p^{-k}} + \delta^{p^{-2k}} v_1^{p^k+p^{-2k}} = 0$, we have

$$\gamma = -\delta v_1^{p^{3k}-p^k} + \delta^{p^{-k}} v_1^{p^{2k}+p^{-k}-p^k-1} - \delta^{p^{-2k}} v_1^{p^{-2k}-1}. \tag{16}$$

From (15) and (16), $\gamma$ and $\delta$ are uniquely determined by $v_1, v_0$ and $\lambda$. Further, there are exactly $p^d - 1$ pairs $(\gamma, \delta)$ corresponding to a given pair $(v_1, v_0)$.

On the other hand, for any $v_0 \in \mathbb{F}_{p^n}^*$ and $\beta \notin \mathbb{F}_{p^d}$, let $v_1 = \beta v_0$. If $\delta$ and $\gamma$ are defined by (15) and (16), respectively, then $\Phi_{\gamma,\delta}(v_1) = 0$. In the sequel, we will prove $v_0 L_{\gamma,\delta}(v_0) = 0$.

From (15), we have

$$\delta v_0^{p^{3k}+1} = \frac{\lambda(\beta^{p^{2k}} - \beta^{p^k})}{(\beta^{2k} - \beta)^{p^k+1}}. \tag{17}$$

Then

$$(\delta v_0^{p^{3k}+1})(\beta - \beta^{p^{2k}}) = \frac{\lambda(\beta^{p^{2k}} - \beta^{p^k})}{(\beta - \beta^{2k})^{p^k}} \text{ and } (\delta v_0^{p^{3k}+1})(\beta^{p^{2k}} - \beta)^{p^k} = \frac{\lambda(\beta^{p^{2k}} - \beta^{p^k})}{\beta^{p^{2k}} - \beta}.$$

Thus, by (16) and (17),

$$
\begin{aligned}
v_0 L_{\gamma,\delta}(v_0) &= \gamma v_0^{p^k+1} + (\gamma v_0^{p^k+1})^{p^{-k}} + \delta v_0^{p^{3k}+1} + (\delta v_0^{p^{3k}+1})^{p^{-3k}} \\
&= \left(-(\delta v_0^{p^{3k}+1})\beta^{p^{3k}-p^k} + (\delta v_0^{p^{3k}+1})^{p^{-k}}\beta^{p^{2k}+p^{-k}-p^k-1} - (\delta v_0^{p^{3k}+1})^{p^{-2k}}\beta^{p^{-2k}-1}\right) + \\
&\quad \left(-(\delta v_0^{p^{3k}+1})^{p^{-k}}\beta^{p^{2k}-1} + (\delta v_0^{p^{3k}+1})^{p^{-2k}}\beta^{p^k+p^{-2k}-1-p^{-k}} - (\delta v_0^{p^{3k}+1})^{p^{-3k}}\beta^{p^{-3k}-p^{-k}}\right) \\
&\quad + \delta v_0^{p^{3k}+1} + (\delta v_0^{p^{3k}+1})^{p^{-3k}} \\
&= (\delta v_0^{p^{3k}+1})(1 - \beta^{p^{3k}-p^k}) + (\delta v_0^{p^{3k}+1})^{p^{-k}}(\beta^{p^{2k}+p^{-k}-p^k-1} - \beta^{p^{2k}-1}) \\
&\quad + (\delta v_0^{p^{3k}+1})^{p^{-2k}}(\beta^{p^k+p^{-2k}-1-p^{-k}} - \beta^{p^{-2k}-1}) + (\delta v_0^{p^{3k}+1})^{p^{-3k}}(1 - \beta^{p^{-3k}-p^{-k}}) \\
&= \beta^{-p^k}(\delta v_0^{p^{3k}+1})(\beta - \beta^{p^{2k}})^{p^k} + \beta^{p^{2k}-p^k-1}(\delta v_0^{p^{3k}+1})^{p^{-k}}(\beta - \beta^{p^{2k}})^{p^{-k}} \\
&\quad + \beta^{p^{-2k}-1-p^{-k}}(\delta v_0^{p^{3k}+1})^{p^{-2k}}(\beta^{p^{2k}} - \beta)^{p^{-k}} + \beta^{-p^{-k}}(\delta v_0^{p^{3k}+1})^{p^{-3k}}(\beta^{p^{2k}} - \beta)^{p^{-3k}} \\
&= \lambda\left(\frac{\beta^{p^{2k}-p^k}-1}{\beta-\beta^{p^{2k}}} + \frac{\beta^{p^{2k}-1}-\beta^{p^{2k}-p^k}}{\beta-\beta^{p^{2k}}} + \frac{\beta^{p^{-2k}-p^{-k}}-\beta^{p^{-2k}-1}}{\beta-\beta^{p^{-2k}}} + \frac{1-\beta^{p^{-2k}-p^{-k}}}{\beta-\beta^{p^{-2k}}}\right) \\
&= \lambda\left(\frac{-\beta^{-1}(\beta-\beta^{p^{2k}})}{\beta-\beta^{p^{2k}}} + \frac{\beta^{-1}(\beta-\beta^{p^{-2k}})}{\beta-\beta^{p^{-2k}}}\right) \\
&= \lambda\left(-\beta^{-1} + \beta^{-1}\right) \\
&= 0.
\end{aligned}
$$

This shows $L_{\gamma,\delta}(v_0) = 0$, and hence $\Phi_{\gamma,\delta}(v_0) = 0$. Thus $\{v_1, v_0\}$ is a basis of the $\mathbb{F}_{p^d}$-vector space consisting of all solutions to $\Phi_{\gamma,\delta}(x) = 0$.

There are totally $\frac{(p^n-1)(p^n-p^d)}{(p^{2d}-1)(p^{2d}-p^d)}$ two-dimensional vector subspaces of $\mathbb{F}_{p^n}$ over $\mathbb{F}_{p^d}$, thus,

$$|R_2| = (p^d - 1) \times \frac{(p^n-1)(p^n-p^d)}{(p^{2d}-1)(p^{2d}-p^d)} = \frac{(p^n-1)(p^{n-d}-1)}{p^{2d}-1}.$$

∎

The values of $S(0, \gamma, \delta)$ can be discussed in terms of $\text{rank}(\Omega_{\gamma,\delta})$ as below.

For $(\gamma, \delta) \in R_0$, $\text{rank}(\Omega_{\gamma,\delta}) = s$ and by a nonsingular linear substitution as in (2), $\Omega_{\gamma,\delta}(x) =$ $\sum_{i=1}^{s} h_i y_i^2$, where $h_i \in \mathbb{F}_{p^d}^*$ and $(y_1, y_2, \cdots, y_s) \in \mathbb{F}_{p^d}^n$. Then

$$
\begin{aligned}
S(0, \gamma, \delta) &= \sum_{x \in \mathbb{F}_{p^n}} \zeta_p^{\mathrm{Tr}_1^d(\Omega_{\gamma,\delta}(x))} \\
&= \sum_{y_1, y_2, \cdots, y_s \in \mathbb{F}_{p^d}} \zeta_p^{\mathrm{Tr}_1^d\left(h_1 y_1^2 + h_2 y_2^2 + \cdots + h_s y_s^2\right)} \\
&= \prod_{i=1}^{s} \sum_{y_i \in \mathbb{F}_{p^d}} \zeta_p^{\mathrm{Tr}_1^d(h_i y_i^2)} \\
&= \begin{cases} \prod_{i=1}^{s} \left( \eta(h_i)(-1)^{d-1} p^{\frac{d}{2}} \right), & p \equiv 1 \pmod{4} \\ \prod_{i=1}^{s} \left( \eta(h_i)(-1)^{d-1}(\sqrt{-1})^d p^{\frac{d}{2}} \right), & p \equiv 3 \pmod{4} \end{cases} \quad \text{by Lemma 1} \\
&= \begin{cases} (-1)^{d-1} \eta(\prod_{i=1}^{s} h_i) p^{\frac{n}{2}}, & p \equiv 1 \pmod{4} \\ (-1)^{d-1} \eta(\prod_{i=1}^{s} h_i)(\sqrt{-1})^n p^{\frac{n}{2}}, & p \equiv 3 \pmod{4}. \end{cases}
\end{aligned}
\tag{18}
$$

Similarly, we have

$$
\begin{aligned}
S(0, \gamma, \delta) &= \sum_{y_1, y_2, \cdots, y_s \in \mathbb{F}_{p^d}} \zeta_p^{\mathrm{Tr}_1^d\left(h_1 y_1^2 + h_2 y_2^2 + \cdots + h_{s-1} y_{s-1}^2\right)} \\
&= p^d \prod_{i=1}^{s-1} \sum_{y_i \in \mathbb{F}_{p^d}} \zeta_p^{\mathrm{Tr}_1^d(h_i y_i^2)} \\
&= \begin{cases} \eta(\prod_{i=1}^{s-1} h_i) p^{\frac{n+d}{2}}, & p \equiv 1 \pmod{4} \\ \eta(\prod_{i=1}^{s-1} h_i)(\sqrt{-1})^{n-d} p^{\frac{n+d}{2}}, & p \equiv 3 \pmod{4} \end{cases}
\end{aligned}
\tag{19}
$$

for $(\gamma, \delta) \in R_1$, and

$$
S(0, \gamma, \delta) = \begin{cases} (-1)^{d-1} \eta(\prod_{i=1}^{s-2} h_i) p^{\frac{n}{2}+d}, & p \equiv 1 \pmod{4} \\ (-1)^{d-1} \eta(\prod_{i=1}^{s-2} h_i)(\sqrt{-1})^{n-2d} p^{\frac{n}{2}+d}, & p \equiv 3 \pmod{4} \end{cases}
\tag{20}
$$

for $(\gamma, \delta) \in R_2$.

From (18), (19) and (20), for $(\gamma, \delta) \in R_i$ with $i \in \{0, 2\}$, we have

$$
S(0, \gamma, \delta) = \sqrt{(-1)^{\frac{p^d-1}{2}}} \theta_i p^{\frac{n+id}{2}}, \quad \theta_i \in \{\pm 1\},
\tag{21}
$$

and for $(\gamma, \delta) \in R_1$,

$$
S(0, \gamma, \delta) = \theta_1 p^{\frac{n+d}{2}}, \quad \theta_1 \in \{\pm 1\}.
\tag{22}
$$

Two subsets $R_{i,j}$ of $R_i$ for $i \in \{0, 1, 2\}$ are defined as

$$
R_{i,j} = \left\{ (\gamma, \delta) \in R_i \mid \theta_i = j \right\}
\tag{23}
$$

where $j = \pm 1$.

The following result can be obtained based on Equalities (18), (20) and the fact that $s$ is odd.

**Lemma 5:** For $i \in \{0, 2\}$, $|R_{i,1}| = |R_{i,-1}|$.

**Proof:** For $i \in \{0, 2\}$, let $(\gamma, \delta) \in R_i$ and $u \in \mathbb{F}_{p^d}^*$ such that $\eta(u) = -1$. Then

$$
\Omega_{u\gamma, u\delta}(x) = \mathrm{Tr}_d^n(u\gamma x^{p^k+1} + u\delta x^{p^{3k}+1}) = u\mathrm{Tr}_d^n(\gamma x^{p^k+1} + \delta x^{p^{3k}+1}) = u\Omega_{\gamma,\delta}(x).
$$

By (18) and (20),

$$S(0, u\gamma, u\delta) = \eta(u)^{s-i} S(0, \gamma, \delta) = (-1)^{s-i} S(0, \gamma, \delta) = -S(0, \gamma, \delta).$$

The above equality shows that for $j \in \{1, -1\}$, if $(\gamma, \delta) \in R_{i,j}$, then $(u\gamma, u\delta) \in R_{i,-j}$. This finishes the proof. ∎

**Proposition 3:** (i) $\sum_{\gamma, \delta \in \mathbb{F}_{p^n}} S(0, \gamma, \delta) = p^{2n}$.

(ii) $\sum_{\gamma, \delta \in \mathbb{F}_{p^n}} S(0, \gamma, \delta)^2 = \begin{cases} p^{2n}(2p^n - 1), & p^d \equiv 1 \pmod 4, \\ p^{2n}, & p^d \equiv 3 \pmod 4. \end{cases}$

**Proof:** The result in (i) can be directly verified, and we only give the proof of (ii).

Notice that

$$\sum_{\gamma,\delta \in \mathbb{F}_{p^n}} S(0,\gamma,\delta)^2$$
$$= \sum_{x,y \in \mathbb{F}_{p^n}} \sum_{\gamma \in \mathbb{F}_{p^n}} \zeta_p^{\mathrm{Tr}_1^n\left(\gamma(x^{p^k+1}+y^{p^k+1})\right)} \sum_{\delta \in \mathbb{F}_{p^n}} \zeta_p^{\mathrm{Tr}_1^n\left(\delta(x^{p^{3k}+1}+y^{p^{3k}+1})\right)}$$
$$= p^{2n}|T_1|,$$

where $T_1$ consists of all solutions $(x, y) \in \mathbb{F}_{p^n} \times \mathbb{F}_{p^n}$ to the equation $x^{p^k+1} + y^{p^k+1} = 0$ since $x^{p^k+1} + y^{p^k+1} = 0$ implies $x^{p^{3k}+1} + y^{p^{3k}+1} = 0$.

If $xy = 0$, $(x, y) = (0, 0)$ is the only solution of $x^{p^k+1} + y^{p^k+1} = 0$.

If $xy \neq 0$, we have $(\frac{x}{y})^{p^k+1} = -1$. If this equation has solution, say $\frac{x}{y} = \alpha^j$ for a primitive element $\alpha$ of $\mathbb{F}_{p^n}$ and $1 \le j < p^n - 1$, then $j(p^k + 1) \equiv \frac{p^n-1}{2} \pmod{p^n - 1}$. This equality holds if and only if $\gcd(p^k+1, p^n-1) | \frac{p^n-1}{2}$. Notice that $\gcd(p^k+1, p^n-1) = 2$ and $s$ is odd. Consequently, $(\frac{x}{y})^{p^k+1} = -1$ has solutions if and only if $p^n \equiv 1 \pmod 4$. Further, in this case the number of solutions is equal to 2. Thus, $x^{p^k+1} + y^{p^k+1} = 0$ has $2(p^n - 1)$ solutions if $p^n \equiv 1 \pmod 4$, and no solution if $p^n \equiv 3 \pmod 4$.

The above analysis and the equality $p^n \equiv p^d \pmod 4$ finish the proof. ∎

With the above preparations, the rank distribution of $\Omega_{\gamma,\delta}(x)$ can be determined as below.

**Proposition 4:** (i) For $i \in \{0, 1, 2\}$ and $j \in \{1, -1\}$, $R_{i,j}$ satisfies

$$\begin{cases} |R_{0,1}| = |R_{0,-1}| = \frac{(p^{n+2d}-p^{n+d}-p^n+p^{2d})(p^n-1)}{2(p^{2d}-1)}, \\ |R_{1,1}| = \frac{(p^{n-d}+p^{\frac{n-d}{2}})(p^n-1)}{2}, \\ |R_{1,-1}| = \frac{(p^{n-d}-p^{\frac{n-d}{2}})(p^n-1)}{2}, \\ |R_{2,1}| = |R_{2,-1}| = \frac{(p^{n-d}-1)(p^n-1)}{2(p^{2d}-1)}. \end{cases}$$

(ii) For odd $s$, when $(\gamma, \delta)$ runs through $\mathbb{F}_{p^n} \times \mathbb{F}_{p^n} \setminus \{(0, 0)\}$, the rank distribution of the quadratic form $\Omega_{\gamma,\delta}(x)$ is given as follows:

$$\begin{cases} s, & \frac{(p^{n+2d}-p^{n+d}-p^n+p^{2d})(p^n-1)}{p^{2d}-1} \text{ times}, \\ s-1, & p^{n-d}(p^n-1) \text{ times}, \\ s-2, & \frac{(p^{n-d}-1)(p^n-1)}{p^{2d}-1} \text{ times}. \end{cases}$$

**Proof:** By Propositions 1, 2, 3, Lemmas 4, and 5, we have the following identities of parameters

$|R_{i,j}|$ with $i \in \{0,1,2\}$ and $j \in \{\pm 1\}$:

$$\begin{cases} |R_0| + |R_1| + |R_2| = p^{2n} - 1, \\ p^{\frac{n+d}{2}}(|R_{1,1}| - |R_{1,-1}|) + p^n = \sum_{\gamma, \delta \in \mathbb{F}_{p^n}} S(0, \gamma, \delta), \\ (-1)^{\frac{p^d-1}{2}} p^n |R_0| + p^{n+d}|R_1| + (-1)^{\frac{p^d-1}{2}} p^{n+2d}|R_2| + p^{2n} = \sum_{\gamma, \delta \in \mathbb{F}_{p^n}} S(0, \gamma, \delta)^2, \\ |R_{0,1}| = |R_{0,-1}|, \\ |R_{2,1}| = |R_{2,-1}| = \frac{(p^{n-d}-1)(p^n-1)}{2(p^{2d}-1)}. \end{cases}$$

This finishes the proof. ∎

By (14), (18)-(23) and Proposition 4, an immediate result is given as below.

**Corollary 2:** For odd $s$, when $(\gamma, \delta)$ runs through $\mathbb{F}_{p^n} \times \mathbb{F}_{p^n} \setminus \{(0,0)\}$, the exponential sum $S(0, \gamma, \delta)$ defined in (3) has the following distribution:

$$\begin{cases} \sqrt{(-1)^{\frac{p^d-1}{2}}} p^{\frac{n}{2}}, & \frac{(p^{n+2d}-p^{n+d}-p^n+p^{2d})(p^n-1)}{2(p^{2d}-1)} \text{ times}, \\ -\sqrt{(-1)^{\frac{p^d-1}{2}}} p^{\frac{n}{2}}, & \frac{(p^{n+2d}-p^{n+d}-p^n+p^{2d})(p^n-1)}{2(p^{2d}-1)} \text{ times}, \\ p^{\frac{n+d}{2}}, & \frac{(p^{n-d}+p^{\frac{n-d}{2}})(p^n-1)}{2} \text{ times}, \\ -p^{\frac{n+d}{2}}, & \frac{(p^{n-d}-p^{\frac{n-d}{2}})(p^n-1)}{2} \text{ times}, \\ \sqrt{(-1)^{\frac{p^d-1}{2}}} p^{\frac{n+2d}{2}}, & \frac{(p^{n-d}-1)(p^n-1)}{2(p^{2d}-1)} \text{ times}, \\ -\sqrt{(-1)^{\frac{p^d-1}{2}}} p^{\frac{n+2d}{2}}, & \frac{(p^{n-d}-1)(p^n-1)}{2(p^{2d}-1)} \text{ times}. \end{cases}$$

## 4 Weight distribution of the $p$-ary code $\mathcal{C}$

This section studies the distribution of the exponential sum $S(\epsilon, \gamma, \delta)$ and the weight distribution of the code $\mathcal{C}$.

If either $\gamma$ or $\delta$ is nonzero, then $\mathrm{Tr}_1^d(\Omega_{\gamma,\delta}(x))$ is also a quadratic form. By (1), Propositions 1, 2 and Corollary 1, $\mathrm{rank}(\mathrm{Tr}_1^d(\Omega_{\gamma,\delta})) = d \cdot \mathrm{rank}(\Omega_{\gamma,\delta}) = n, n-d$, or $n-2d$.

For $\rho \in \mathbb{F}_p$, let $N_{\epsilon,\gamma,\delta}(\rho)$ denote the number of solutions to $\mathrm{Tr}_1^d(\Omega_{\gamma,\delta}(x)) + \mathrm{Tr}_1^n(\epsilon x) = \rho$. Then, (3) can be written as

$$S(\epsilon, \gamma, \delta) = \sum_{\rho=0}^{p-1} N_{\epsilon,\gamma,\delta}(\rho) \zeta_p^\rho. \tag{24}$$

Let $\{\alpha_1, \alpha_2, \cdots, \alpha_n\}$ be a basis of $\mathbb{F}_{p^n}$ over $\mathbb{F}_p$, and $\epsilon = \sum_{i=1}^n \epsilon_i \alpha_i$ with $\epsilon_i \in \mathbb{F}_p$. Then the matrix $C = (\mathrm{Tr}_1^n(\alpha_i \alpha_j))_{1 \le i,j \le n}$ is nonsingular. Let $D^\mathrm{T} = (\epsilon_1, \epsilon_2, \cdots, \epsilon_n) \in \mathbb{F}_p^n$ and $X = BY$ be defined as in Section 2, then $\mathrm{Tr}_1^n(\epsilon x) = D^\mathrm{T} C X$. Denote $D^\mathrm{T} C B = (b_1, b_2, \cdots, b_n)$, and we have

$$\mathrm{Tr}_1^d(\Omega_{\gamma,\delta}(x)) + \mathrm{Tr}_1^n(\epsilon x) = Y^\mathrm{T} B^\mathrm{T} A B Y + D^\mathrm{T} C B Y = \sum_{i=1}^n a_i y_i^2 + \sum_{i=1}^n b_i y_i. \tag{25}$$

By application of the quadratic form theory, the distribution of $S(\epsilon, \gamma, \delta)$ is discussed and the weight distribution of $\mathcal{C}$ is determined.

**Theorem 1:** For two positive integers $n$ and $k$ with $d = \gcd(n, k)$, if $s$ is odd, then when $(\epsilon, \gamma, \delta)$

runs through $\mathbb{F}_{p^n} \times \mathbb{F}_{p^n} \times \mathbb{F}_{p^n}$, the exponential sum $S(\epsilon, \gamma, \delta)$ defined in (3) has the following distribution

$$\begin{cases} p^n, & 1 \text{ time,} \\ 0, & (p^n-1)(p^{2n-d}-p^{2n-2d}+p^{2n-3d}-p^{n-2d}+1) \text{ times,} \\ \sqrt{(-1)^{\frac{p-1}{2}}}p^{\frac{n}{2}}\zeta_p^\rho, & \frac{(p^{n-1}+\eta(-\rho)p^{\frac{n-1}{2}})(p^{n+2d}-p^{n+d}-p^n+p^{2d})(p^n-1)}{2(p^{2d}-1)} \text{ times,} \\ -\sqrt{(-1)^{\frac{p-1}{2}}}p^{\frac{n}{2}}\zeta_p^\rho, & \frac{(p^{n-1}-\eta(-\rho)p^{\frac{n-1}{2}})(p^{n+2d}-p^{n+d}-p^n+p^{2d})(p^n-1)}{2(p^{2d}-1)} \text{ times,} \\ p^{\frac{n+d}{2}}\zeta_p^\rho, & \frac{(p^{n-d-1}+\upsilon(\rho)p^{\frac{n-d-2}{2}})(p^{n-d}+p^{\frac{n-d}{2}})(p^n-1)}{2} \text{ times,} \\ -p^{\frac{n+d}{2}}\zeta_p^\rho, & \frac{(p^{n-d-1}-\upsilon(\rho)p^{\frac{n-d-2}{2}})(p^{n-d}-p^{\frac{n-d}{2}})(p^n-1)}{2} \text{ times,} \\ \sqrt{(-1)^{\frac{p-1}{2}}}p^{\frac{n+2d}{2}}\zeta_p^\rho, & \frac{(p^{n-2d-1}+\eta(-\rho)p^{\frac{n-2d-1}{2}})(p^{n-d}-1)(p^n-1)}{2(p^{2d}-1)} \text{ times,} \\ -\sqrt{(-1)^{\frac{p-1}{2}}}p^{\frac{n+2d}{2}}\zeta_p^\rho, & \frac{(p^{n-2d-1}-\eta(-\rho)p^{\frac{n-2d-1}{2}})(p^{n-d}-1)(p^n-1)}{2(p^{2d}-1)} \text{ times} \end{cases}$$

for odd $d$, and

$$\begin{cases} p^n, & 1 \text{ time,} \\ 0, & (p^n-1)(p^{2n-d}-p^{2n-2d}+p^{2n-3d}-p^{n-2d}+1) \text{ times,} \\ p^{\frac{n}{2}}\zeta_p^\rho, & \frac{(p^{n-1}+\upsilon(\rho)p^{\frac{n-2}{2}})(p^{n+2d}-p^{n+d}-p^n+p^{2d})(p^n-1)}{2(p^{2d}-1)} \text{ times,} \\ -p^{\frac{n}{2}}\zeta_p^\rho, & \frac{(p^{n-1}-\upsilon(\rho)p^{\frac{n-2}{2}})(p^{n+2d}-p^{n+d}-p^n+p^{2d})(p^n-1)}{2(p^{2d}-1)} \text{ times,} \\ p^{\frac{n+d}{2}}\zeta_p^\rho, & \frac{(p^{n-d-1}+\upsilon(\rho)p^{\frac{n-d-2}{2}})(p^{n-d}+p^{\frac{n-d}{2}})(p^n-1)}{2} \text{ times,} \\ -p^{\frac{n+d}{2}}\zeta_p^\rho, & \frac{(p^{n-d-1}-\upsilon(\rho)p^{\frac{n-d-2}{2}})(p^{n-d}-p^{\frac{n-d}{2}})(p^n-1)}{2} \text{ times,} \\ p^{\frac{n+2d}{2}}\zeta_p^\rho, & \frac{(p^{n-2d-1}+\upsilon(\rho)p^{\frac{n-2d-2}{2}})(p^{n-d}-1)(p^n-1)}{2(p^{2d}-1)} \text{ times,} \\ -p^{\frac{n+2d}{2}}\zeta_p^\rho, & \frac{(p^{n-2d-1}-\upsilon(\rho)p^{\frac{n-2d-2}{2}})(p^{n-d}-1)(p^n-1)}{2(p^{2d}-1)} \text{ times} \end{cases}$$

for even $d$, where $\rho = 0, 1, \cdots, p-1$, $\eta$ is the quadratic character of $\mathbb{F}_p$ and $\upsilon(0) = p-1$, $\upsilon(\rho) = -1$ for $\rho \in \mathbb{F}_p^*$.

**Proof:** Since $s$ is odd, the integer $n-d$ is always even. If $d$ is odd, then $n$ and $n-2d$ are both odd. The proof in this case is divided into the following subcases.

(i) For $(\gamma, \delta) = (0,0)$, $S(\epsilon, 0, 0) = 0$ for $\epsilon \neq 0$, and $p^n$ for $\epsilon = 0$.

(ii) For $(\gamma, \delta) \neq (0,0)$, the discussion is divided into three subcases.

In the case of $(\gamma, \delta) \in R_0$, for $1 \leq i \leq n$, let $y_i = z_i - \frac{b_i}{2a_i}$. Then $\sum_{i=1}^{n}(a_i y_i^2 + b_i y_i) = \rho$ is equivalent to $\sum_{i=1}^{n} a_i z_i^2 = \lambda_{\epsilon, \gamma, \delta} + \rho$, where $\lambda_{\epsilon, \gamma, \delta} = \sum_{i=1}^{n} \frac{b_i^2}{4a_i}$. Let $\Delta_0 = \prod_{i=1}^{n} a_i$, then Lemma 2 implies

$$N_{\epsilon, \gamma, \delta}(\rho) = p^{n-1} + p^{\frac{n-1}{2}}\eta\left((-1)^{\frac{n-1}{2}}(\lambda_{\epsilon,\gamma,\delta}+\rho)\Delta_0\right). \tag{26}$$

Notice that the matrix $CB$ in (25) is nonsingular. As a consequence, $(b_1, b_2, \cdots, b_n)$ runs through $\mathbb{F}_p^n$ as $\epsilon$ runs through $\mathbb{F}_{p^n}$. $\lambda_{\epsilon,\gamma,\delta}$ is also a quadratic form with $n$ variables $b_i$ for $1 \leq i \leq n$. Again by Lemma 2, as $\epsilon$ runs through $\mathbb{F}_{p^n}$,

$$\lambda_{\epsilon,\gamma,\delta} = \sum_{i=1}^{n} \frac{b_i^2}{4a_i} = \rho' \text{ occurring } p^{n-1} + p^{\frac{n-1}{2}}\eta\left((-1)^{\frac{n-1}{2}}\rho'\Delta_0\right) \text{ times} \tag{27}$$

for each $\rho' \in \mathbb{F}_p$ since $\eta\left((4^n \prod_{i=1}^{n} a_i)^{-1}\right) = \eta(\prod_{i=1}^{n} a_i)$.

By (24), (26) and Lemma 3 (i), we have

$$S(\epsilon, \gamma, \delta) = \eta\left((-1)^{\frac{n-1}{2}}\Delta_0\right)p^{\frac{n}{2}}\sqrt{(-1)^{\frac{p-1}{2}}}\zeta_p^{-\lambda_{\epsilon,\gamma,\delta}}. \tag{28}$$

By (27), as $\epsilon$ runs through $\mathbb{F}_{p^n}$, for each $\rho \in \mathbb{F}_p$, we have

$$S(\epsilon, \gamma, \delta) = \eta\left((-1)^{\frac{n-1}{2}}\Delta_0\right)\sqrt{(-1)^{\frac{p-1}{2}}p}^{\frac{n}{2}}\zeta_p^\rho \text{ occuring } p^{n-1} + p^{\frac{n-1}{2}}\eta\left((-1)^{\frac{n+1}{2}}\rho\Delta_0\right) \text{ times.} \quad (29)$$

In the case of $(\gamma, \delta) \in R_1$, the rank of $\mathrm{Tr}_1^d(\Omega_{\gamma,\delta}(x))$ is $n-d$, and then

$$\mathrm{Tr}_1^d(\Omega_{\gamma,\delta}(x)) + \mathrm{Tr}_1^n(\epsilon x) = \sum_{i=1}^{n-d} a_i y_i^2 + \sum_{i=1}^{n} b_i y_i.$$

If there exists some $b_i \neq 0$ for $n-d < i \leq n$, then for any $\rho \in \mathbb{F}_p$, $N_{\epsilon,\gamma,\delta}(\rho) = p^{n-1}$ and $S(\epsilon, \gamma, \delta) = 0$. Since the matrix $CB$ is nonsingular, there are exactly $p^n - p^{n-d}$ choices for $\epsilon$ such that there is at least one $b_i \neq 0$ with $n-d < i \leq n$, as $\epsilon$ runs through $\mathbb{F}_{p^n}$.

If $b_i = 0$ for all $n-d < i \leq n$, then $\sum_{i=1}^{n-d}(a_i y_i^2 + b_i y_i) = \rho$ is equivalent to $\sum_{i=1}^{n-d} a_i z_i^2 = \lambda_{\epsilon,\gamma,\delta} + \rho$, where $\lambda_{\epsilon,\gamma,\delta} = \sum_{i=1}^{n-d}\frac{b_i^2}{4a_i}$ and $z_i = y_i + \frac{b_i}{2a_i}$ for $1 \leq i \leq n-d$. Let $\Delta_1 = \prod_{i=1}^{n-d} a_i$, then for any $\rho \in \mathbb{F}_p$ and even $n-d$, by Lemma 2,

$$N_{\epsilon,\gamma,\delta}(\rho) = p^d\left(p^{n-d-1} + v(\lambda_{\epsilon,\gamma,\delta} + \rho)p^{\frac{n-d-2}{2}}\eta((-1)^{\frac{n-d}{2}}\Delta_1)\right),$$

i.e.,

$$N_{\epsilon,\gamma,\delta}(\rho) = p^{n-1} + v(\lambda_{\epsilon,\gamma,\delta} + \rho)p^{\frac{n+d-2}{2}}\eta\left((-1)^{\frac{n-d}{2}}\Delta_1\right). \quad (30)$$

By Lemma 2, when $(b_1, b_2, \cdots, b_{n-d})$ runs through $\mathbb{F}_p^{n-d}$,

$$\lambda_{\epsilon,\gamma,\delta} = \sum_{i=1}^{n-d}\frac{b_i^2}{4a_i} = \rho' \text{ occurring } p^{n-d-1} + v(\rho')p^{\frac{n-d-2}{2}}\eta\left((-1)^{\frac{n-d}{2}}\Delta_1\right) \text{ times} \quad (31)$$

for each $\rho' \in \mathbb{F}_p$. Then by (24) and (30),

$$S(\epsilon, \gamma, \delta) = \eta\left((-1)^{\frac{n-d}{2}}\Delta_1\right)p^{\frac{n+d}{2}}\zeta_p^{-\lambda_{\epsilon,\gamma,\delta}}$$

since $\sum_{\rho \in \mathbb{F}_p} v(\rho + \lambda_{\gamma,\delta,\epsilon})\zeta_p^{\rho+\lambda_{\gamma,\delta,\epsilon}} = p$ by Lemma 3 (ii). Notice that $v(-\rho) = v(\rho)$ for any $\rho \in \mathbb{F}_p$. By (31), when $(b_1, b_2, \cdots, b_{n-d})$ runs through $\mathbb{F}_p^{n-d}$,

$$S(\epsilon, \gamma, \delta) = \eta\left((-1)^{\frac{n-d}{2}}\Delta_1\right)p^{\frac{n+d}{2}}\zeta_p^\rho \text{ occuring } p^{n-d-1} + v(\rho)p^{\frac{n-d-2}{2}}\eta\left((-1)^{\frac{n-d}{2}}\Delta_1\right) \text{ times} \quad (32)$$

for each $\rho \in \mathbb{F}_p$.

In the case of $(\gamma, \delta) \in R_2$, the rank of $\mathrm{Tr}_1^d(\Omega_{\gamma,\delta}(x))$ is $n - 2d$ and

$$\mathrm{Tr}_1^d(\Omega_{\gamma,\delta}(x)) + \mathrm{Tr}_1^n(\epsilon x) = \sum_{i=1}^{n-2d} a_i y_i^2 + \sum_{i=1}^{n} b_i y_i.$$

Similarly, if there exists some $b_i \neq 0$ with $n - 2d < i \leq n$, then $N_{\epsilon,\gamma,\delta}(\rho) = p^{n-1}$ for any $\rho \in \mathbb{F}_p$ and $S(\epsilon, \gamma, \delta) = 0$. When $\epsilon$ runs through $\mathbb{F}_{p^n}$, there are $p^n - p^{n-2d}$ choices for $\epsilon$ such that there is at least one $b_i \neq 0$ with $n - 2d < i \leq n$.

If $b_i = 0$ for all $n - 2d < i \leq n$, a similar analysis shows that for any $\rho \in \mathbb{F}_p$, by Lemma 2,

$$N_{\epsilon,\gamma,\delta}(\rho) = p^{n-1} + p^{\frac{n+2d-1}{2}}\eta\left((-1)^{\frac{n-2d-1}{2}}(\lambda_{\epsilon,\gamma,\delta} + \rho)\Delta_2\right) \quad (33)$$

where $\lambda_{\epsilon,\gamma,\delta} = \sum_{i=1}^{n-2d}\frac{b_i^2}{4a_i}$ and $\Delta_2 = \prod_{i=1}^{n-2d} a_i$. When $(b_1, b_2, \cdots, b_{n-2d})$ runs through $\mathbb{F}_p^{n-2d}$, by Lemma 2,

$$\lambda_{\epsilon,\gamma,\delta} = \sum_{i=1}^{n-2d}\frac{b_i^2}{4a_i} = \rho' \text{ occuring } p^{n-2d-1} + p^{\frac{n-2d-1}{2}}\eta\left((-1)^{\frac{n-2d-1}{2}}\rho'\Delta_2\right) \text{ times} \quad (34)$$

for each $\rho' \in \mathbb{F}_p$. Thus, by lemma 3 (i), (24) and (33), we have

$$S(\gamma, \delta, \epsilon) = \eta\left((-1)^{\frac{n-2d-1}{2}}\Delta_2\right)\sqrt{(-1)^{\frac{p-1}{2}}}p^{\frac{n}{2}+d}\zeta_p^{-\lambda_{\gamma,\delta,\epsilon}}.$$

Consequently, when $(b_1, b_2, \cdots, b_{n-2d})$ runs through $\mathbb{F}_p^{n-2d}$,

$$S(\epsilon, \gamma, \delta) = \eta\left((-1)^{\frac{n-2d-1}{2}}\Delta_2\right)\sqrt{(-1)^{\frac{p-1}{2}}}p^{\frac{n}{2}+d}\zeta_p^{\rho} \text{ occuring} \\ p^{n-2d-1} + p^{\frac{n-2d-1}{2}}\eta\left((-1)^{\frac{n-2d+1}{2}}\rho\Delta_2\right) \text{ times} \qquad (35)$$

for each $\rho \in \mathbb{F}_p$.

From the above analysis, $S(\epsilon, \gamma, \delta) = p^n$ if and only if $\epsilon = \gamma = \delta = 0$, and $S(\epsilon, \gamma, \delta) = 0$ occurs $p^n - 1 + (p^n - p^{n-d})|R_1| + (p^n - p^{n-2d})|R_2| = (p^n - 1)(p^{2n-d} - p^{2n-2d} + p^{2n-3d} - p^{n-2d} + 1)$ times. By (28) and Corollary 2, for $i \in \{1, -1\}$, there are $|R_{0,i}|$ pairs $(\gamma, \delta) \in \mathbb{F}_{p^n} \times \mathbb{F}_{p^n}$ such that $\eta\left((-1)^{\frac{n-1}{2}}\Delta_0\right) = i$. Thus for each $\rho \in \mathbb{F}_p$, we have

$$S(\epsilon, \gamma, \delta) = \pm\sqrt{(-1)^{\frac{p-1}{2}}}p^{\frac{n}{2}}\zeta_p^{\rho} \text{ occuring } \left(p^{n-1} \pm p^{\frac{n-1}{2}}\eta(-\rho)\right)|R_{0,\pm 1}| \text{ times}$$

when $(\epsilon, \gamma, \delta)$ runs through $\mathbb{F}_{p^n} \times \mathbb{F}_{p^n} \times \mathbb{F}_{p^n}$. The other cases can be similarly analyzed.

For the even case of $d$, the integers $n$, $n - 2d$ are also even. This case has a difference from the odd case of $d$ only in the application of Lemma 2. It can be proven in a similar way and we omit the proof here. ∎

Notice that the weight of the codeword $\mathbf{c}(\epsilon, \gamma, \delta)$ is equal to $p^n - 1 - (N_{\epsilon,\gamma,\delta}(0) - 1) = p^n - N_{\epsilon,\gamma,\delta}(0)$. Consequently, the values $N_{\epsilon,\gamma,\delta}(0)$ for any given $\epsilon, \gamma, \delta$ are needed to determine the weight distribution.

**Theorem 2:** For two integers $n$ and $k$ with $d = \gcd(n, k)$, if $s = n/d$ is odd, then the weight distribution of the code $\mathcal{C}$ is given by

$$\begin{cases}
0, & 1 \text{ time,} \\
(p-1)p^{n-1}, & (p^n-1)\left(p^{2n-1} + (p-1)p^{2n-d-1} - p^{2n-2d} + \right. \\
& \left. (p-1)p^{2n-3d-1} + p^{n-1} - (p-1)p^{n-2d-1} + 1\right) \text{ times,} \\
(p-1)p^{n-1} - p^{\frac{n-1}{2}}, & \frac{(p-1)(p^{n-1}+p^{\frac{n-1}{2}})(p^{n+2d}-p^{n+d}-p^n+p^{2d})(p^n-1)}{2(p^{2d}-1)} \text{ times,} \\
(p-1)p^{n-1} + p^{\frac{n-1}{2}}, & \frac{(p-1)(p^{n-1}-p^{\frac{n-1}{2}})(p^{n+2d}-p^{n+d}-p^n+p^{2d})(p^n-1)}{2(p^{2d}-1)} \text{ times,} \\
(p-1)(p^{n-1} - p^{\frac{n+d-2}{2}}), & \frac{(p^{n-d-1}+(p-1)p^{\frac{n-d-2}{2}})(p^{n-d}+p^{\frac{n-d}{2}})(p^n-1)}{2} \text{ times,} \\
(p-1)(p^{n-1} + p^{\frac{n+d-2}{2}}), & \frac{(p^{n-d-1}-(p-1)p^{\frac{n-d-2}{2}})(p^{n-d}-p^{\frac{n-d}{2}})(p^n-1)}{2} \text{ times,} \\
(p-1)p^{n-1} - p^{\frac{n+d-2}{2}}, & \frac{(p-1)(p^{n-d-1}+p^{\frac{n-d-2}{2}})(p^{n-d}-p^{\frac{n-d}{2}})(p^n-1)}{2} \text{ times,} \\
(p-1)p^{n-1} + p^{\frac{n+d-2}{2}}, & \frac{(p-1)(p^{n-d-1}-p^{\frac{n-d-2}{2}})(p^{n-d}+p^{\frac{n-d}{2}})(p^n-1)}{2} \text{ times,} \\
(p-1)p^{n-1} - p^{\frac{n+2d-1}{2}}, & \frac{(p-1)(p^{n-2d-1}+p^{\frac{n-2d-1}{2}})(p^{n-d}-1)(p^n-1)}{2(p^{2d}-1)} \text{ times,} \\
(p-1)p^{n-1} + p^{\frac{n+2d-1}{2}}, & \frac{(p-1)(p^{n-2d-1}-p^{\frac{n-2d-1}{2}})(p^{n-d}-1)(p^n-1)}{2(p^{2d}-1)} \text{ times}
\end{cases}$$

for odd $d$, and

$$\begin{cases} 0, & 1 \text{ time,} \\ (p-1)p^{n-1}, & (p^n-1)\left(p^{2n-d}-p^{2n-2d}+p^{2n-3d}-p^{n-2d}+1\right) \text{ times,} \\ (p-1)(p^{n-1}-p^{\frac{n-2}{2}}), & \frac{(p^{n-1}+(p-1)p^{\frac{n-2}{2}})(p^{n+2d}-p^{n+d}-p^n+p^{2d})(p^n-1)}{2(p^{2d}-1)} \text{ times,} \\ (p-1)(p^{n-1}+p^{\frac{n-2}{2}}), & \frac{(p^{n-1}-(p-1)p^{\frac{n-2}{2}})(p^{n+2d}-p^{n+d}-p^n+p^{2d})(p^n-1)}{2(p^{2d}-1)} \text{ times,} \\ (p-1)p^{n-1}-p^{\frac{n-2}{2}}, & \frac{(p-1)(p^{n-1}+p^{\frac{n-2}{2}})(p^{n+2d}-p^{n+d}-p^n+p^{2d})(p^n-1)}{2(p^{2d}-1)} \text{ times,} \\ (p-1)p^{n-1}+p^{\frac{n-2}{2}}, & \frac{(p-1)(p^{n-1}-p^{\frac{n-2}{2}})(p^{n+2d}-p^{n+d}-p^n+p^{2d})(p^n-1)}{2(p^{2d}-1)} \text{ times,} \\ (p-1)(p^{n-1}-p^{\frac{n+d-2}{2}}), & \frac{(p^{n-d-1}+(p-1)p^{\frac{n-d-2}{2}})(p^{n-d}+p^{\frac{n-d}{2}})(p^n-1)}{2} \text{ times,} \\ (p-1)(p^{n-1}+p^{\frac{n+d-2}{2}}), & \frac{(p^{n-d-1}-(p-1)p^{\frac{n-d-2}{2}})(p^{n-d}-p^{\frac{n-d}{2}})(p^n-1)}{2} \text{ times,} \\ (p-1)p^{n-1}-p^{\frac{n+d-2}{2}}, & \frac{(p-1)(p^{n-d-1}+p^{\frac{n-d-2}{2}})(p^{n-d}-p^{\frac{n-d}{2}})(p^n-1)}{2} \text{ times,} \\ (p-1)p^{n-1}+p^{\frac{n+d-2}{2}}, & \frac{(p-1)(p^{n-d-1}-p^{\frac{n-d-2}{2}})(p^{n-d}+p^{\frac{n-d}{2}})(p^n-1)}{2} \text{ times,} \\ (p-1)(p^{n-1}-p^{\frac{n+2d-2}{2}}), & \frac{(p^{n-2d-1}+(p-1)p^{\frac{n-2d-2}{2}})(p^{n-d}-1)(p^n-1)}{2(p^{2d}-1)} \text{ times,} \\ (p-1)(p^{n-1}+p^{\frac{n+2d-2}{2}}), & \frac{(p^{n-2d-1}-(p-1)p^{\frac{n-2d-2}{2}})(p^{n-d}-1)(p^n-1)}{2(p^{2d}-1)} \text{ times,} \\ (p-1)p^{n-1}-p^{\frac{n+2d-2}{2}}, & \frac{(p-1)(p^{n-2d-1}+p^{\frac{n-2d-2}{2}})(p^{n-d}-1)(p^n-1)}{2(p^{2d}-1)} \text{ times,} \\ (p-1)p^{n-1}+p^{\frac{n+2d-2}{2}}, & \frac{(p-1)(p^{n-2d-1}-p^{\frac{n-2d-2}{2}})(p^{n-d}-1)(p^n-1)}{2(p^{2d}-1)} \text{ times.} \end{cases}$$

for even $d$, as $(\epsilon, \gamma, \delta)$ runs through $\mathbb{F}_{p^n} \times \mathbb{F}_{p^n} \times \mathbb{F}_{p^n}$.

**Proof:** We also only give the proof for odd $d$, and omit the proof of the other case.

(i) For $(\gamma, \delta) = (0,0)$, $N_{\epsilon, \gamma, \delta}(0) = p^{n-1}$ for $\epsilon \neq 0$, and $p^n$ for $\epsilon = 0$.

(ii) For $(\gamma, \delta) \in R_0$. Notice that there are $\frac{p-1}{2}$ square and non-square elements in $\mathbb{F}_p^*$, respectively. As $\epsilon$ runs through $\mathbb{F}_{p^n}$, by (26) and (27),

$$N_{\epsilon, \gamma, \delta}(0) = p^{n-1} \quad \text{occuring} \quad p^{n-1} \text{ times}$$

and

$$N_{\epsilon, \gamma, \delta}(0) = p^{n-1} \pm p^{\frac{n-1}{2}} \eta\left((-1)^{\frac{n-1}{2}} \Delta_0\right) \quad \text{occuring} \quad \frac{p-1}{2}\left(p^{n-1} \pm p^{\frac{n-1}{2}} \eta((-1)^{\frac{n-1}{2}} \Delta_0)\right) \text{ times.}$$

For $(\gamma, \delta) \in R_1$, if there exists some $b_i \neq 0$ for $n-d < i \leq n$, then for any $\rho \in \mathbb{F}_p$, $N_{\epsilon, \gamma, \delta}(\rho) = p^{n-1}$. If $b_i = 0$ for all $n-d < i \leq n$, when $(b_1, b_2, \cdots, b_{n-d})$ runs through $\mathbb{F}_p^{n-d}$,

$$N_{\epsilon, \gamma, \delta}(0) = p^{n-1} + (p-1)p^{\frac{n+d-2}{2}} \eta\left((-1)^{\frac{n-d}{2}} \Delta_1\right) \text{ occuring } p^{n-d-1} + (p-1)p^{\frac{n-d-2}{2}} \eta\left((-1)^{\frac{n-d}{2}} \Delta_1\right) \text{ times}$$

and

$$N_{\epsilon, \gamma, \delta}(0) = p^{n-1} - p^{\frac{n+d-2}{2}} \eta\left((-1)^{\frac{n-d}{2}} \Delta_1\right) \text{ occuring } (p-1)\left(p^{n-d-1} - p^{\frac{n-d-2}{2}} \eta((-1)^{\frac{n-d}{2}} \Delta_1)\right) \text{ times.}$$

For $(\gamma, \delta) \in R_2$, if there exists some $b_i \neq 0$ with $n - 2d < i \leq n$, then $N_{\epsilon, \gamma, \delta}(\rho) = p^{n-1}$ for any $\rho \in \mathbb{F}_p$.

If $b_i = 0$ for all $n - 2d < i \leq n$, when $(b_1, b_2, \cdots, b_{n-2d})$ runs through $\mathbb{F}_p^{n-2d}$,

$$N_{\gamma, \delta, \epsilon}(0) = p^{n-1} \quad \text{occuring} \quad p^{n-2d-1} \text{ times,}$$

and

$$N_{\gamma, \delta, \epsilon}(0) = p^{n-1} \pm p^{\frac{n+2d-1}{2}} \eta\left((-1)^{\frac{n-2d-1}{2}} \Delta_2\right) \text{ occuring}$$
$$\frac{p-1}{2}\left(p^{n-2d-1} \pm p^{\frac{n-2d-1}{2}} \eta((-1)^{\frac{n-2d-1}{2}} \Delta_2)\right) \text{ times.}$$

We only give the frequencies of the codewords with weight $(p-1)p^{n-1}$ and $(p-1)p^{n-1} - p^{\frac{n-1}{2}}$. Other cases can be similarly analyzed. The weight of $\mathbf{c}(\epsilon, \gamma, \delta)$ is equal to $(p-1)p^{n-1}$ if and only if $N_{\epsilon,\gamma,\delta}(0) = p^{n-1}$. By the above analysis and Proposition 4, the frequency is equal to

$$
\begin{aligned}
&p^n - 1 + p^{n-1}|R_0| + \left(p^n - p^{n-d}\right)|R_1| + (p^n - p^{n-2d} + p^{n-2d-1})|R_2| \\
&= (p^n - 1)\left(p^{2n-1} + (p-1)p^{2n-d-1} - p^{2n-2d} + (p-1)p^{2n-3d-1} + p^{n-1} - (p-1)p^{n-2d-1} + 1\right).
\end{aligned}
$$

The weight of $\mathbf{c}(\epsilon, \gamma, \delta)$ is equal to $(p-1)p^{n-1} - p^{\frac{n-1}{2}}$ if and only if $N_{\epsilon,\gamma,\delta}(0) = p^{n-1} + p^{\frac{n-1}{2}}$. The corresponding frequency is

$$
\begin{aligned}
&\frac{p-1}{2}(p^{n-1} + p^{\frac{n-1}{2}})|R_{0,1}| + \frac{p-1}{2}(p^{n-1} + p^{\frac{n-1}{2}})|R_{0,-1}| \\
&= \frac{(p-1)(p^{n-1} + p^{\frac{n-1}{2}})(p^{n+2d} - p^{n+d} - p^n + p^{2d})(p^n - 1)}{2(p^{2d} - 1)}.
\end{aligned}
$$

∎

# Acknowledgments

This research was partially supported by the National Natural Science Foundation of China under Grants 60603012 and 60573053.